\documentclass[11pt, pre, notitlepage]{revtex4-1}
\usepackage{graphicx}
\usepackage[space]{grffile}
\usepackage{latexsym}
\usepackage{textcomp}
\usepackage{longtable}
\usepackage{tabulary}
\usepackage{booktabs,array,multirow}
\usepackage{amsfonts,amsmath,amssymb}
\usepackage{natbib}
\usepackage{url}
\usepackage{hyperref}
\hypersetup{colorlinks=false,pdfborder={0 0 0}}
\usepackage{etoolbox}
\makeatletter

\usepackage[utf8]{inputenc}

\begin{document}

\title{Intermittency as metastability: a predictive approach to evolution in
disordered environments}

\author{Matteo Smerlak}
\affiliation{Max Planck Institute for Mathematics in the Sciences, Leipzig, Germany}

% Please give the surname of the lead author for the running footer
%\author{Matteo Smerlak} 

% Please add here a significance statement to explain the relevance of your work

% Keywords are not mandatory, but authors are strongly encouraged to provide them. If provided, please include two to five keywords, separated by the pipe symbol, e.g:

\begin{abstract}
Many systems across the sciences evolve through a combination of
multiplicative growth and diffusive transport. In the presence of
disorder, these systems tend to form localized structures which
alternate between long periods of relative stasis and short bursts of
activity. This behaviour, known as~\emph{intermittency}~in physics
and~\emph{punctuated equilibrium~}in evolutionary theory, is difficult
to forecast; in particular there is no general principle to locate the
regions where the system will settle, how long it will stay there, or
where it will jump next. Here I introduce a predictive theory of linear
intermittency that closes these gaps. I show that any positive linear
system can be mapped onto a generalization of the ``maximal entropy
random walk'', a Markov process on graphs with non-local transition
rates. This construction reveals the localization islands as local
minima of an effective potential, and intermittent jumps as barrier
crossings in that potential. My results unify the concepts of
linear intermittency and Markovian metastability, and provide a generally applicable
method to reduce, and predict, the dynamics of evolutionary processes. Applications span physics, evolutionary dynamics and epidemiology.%
\end{abstract}%

\maketitle

\section{Introduction}
Reaction and diffusion, selection and mutation, infection and transmission: many evolutionary processes across the sciences are powered by the interaction of multiplicative growth and transport. An intriguing but common feature of these systems is their proneness to \textit{intermittency}: instead of spreading smoothly through their state space, they tend to form long-lived, localized structures which eventually destabilize in short bursts of seemingly unpredictable activity. In the context of Darwinian evolution, this behavior is known as ``punctuated equilibrium'' and is observed in the fossil record \cite{eldredge1977} as well as in experiments with viruses \cite{chao1999} and bacteria \cite{Lenski_1994}. In epidemiology, localized clusters of infection are ``hotspots'' \cite{Lessler_2017}. In the context of human development, localized structures are nothing but cities \cite{Zanette_1997}. In early cosmology, they are large scale structures \cite{Shandarin_1989}. Examples abound. 

One reason for the prevalence of intermittency in growth-transport processes is their sensitivity to disorder. Unlike additive noise, which blurs an image without changing its contours, small variations in growth rates can build up to macroscopic effects over time: no matter how small its coefficient, an exponential with rate $r$ will always overtake another exponential with rate $r'$ if $r>r'$. Thence follows the basic mechanism of intermittency in random media \cite{Zel_dovich_1987}: a disordered environment provides seeds (or ``islands'') where the field can grow at faster-than-average rates; transport, although too weak to spread such self-amplifying, localized structures, permits the discovery of new seeds in remote regions of the state space; if the new structures have a higher growth rate than the old ones, they will necessarily outgrow them in the long run, and the system will seem to ''jump'' from one island to another. Note that this scenario does not require involves any non-linearity---structures form purely through the conflicting effects of linear amplification and relaxation mechanisms.  

\begin{figure}[t!]
\begin{center}
\includegraphics[width=\columnwidth]{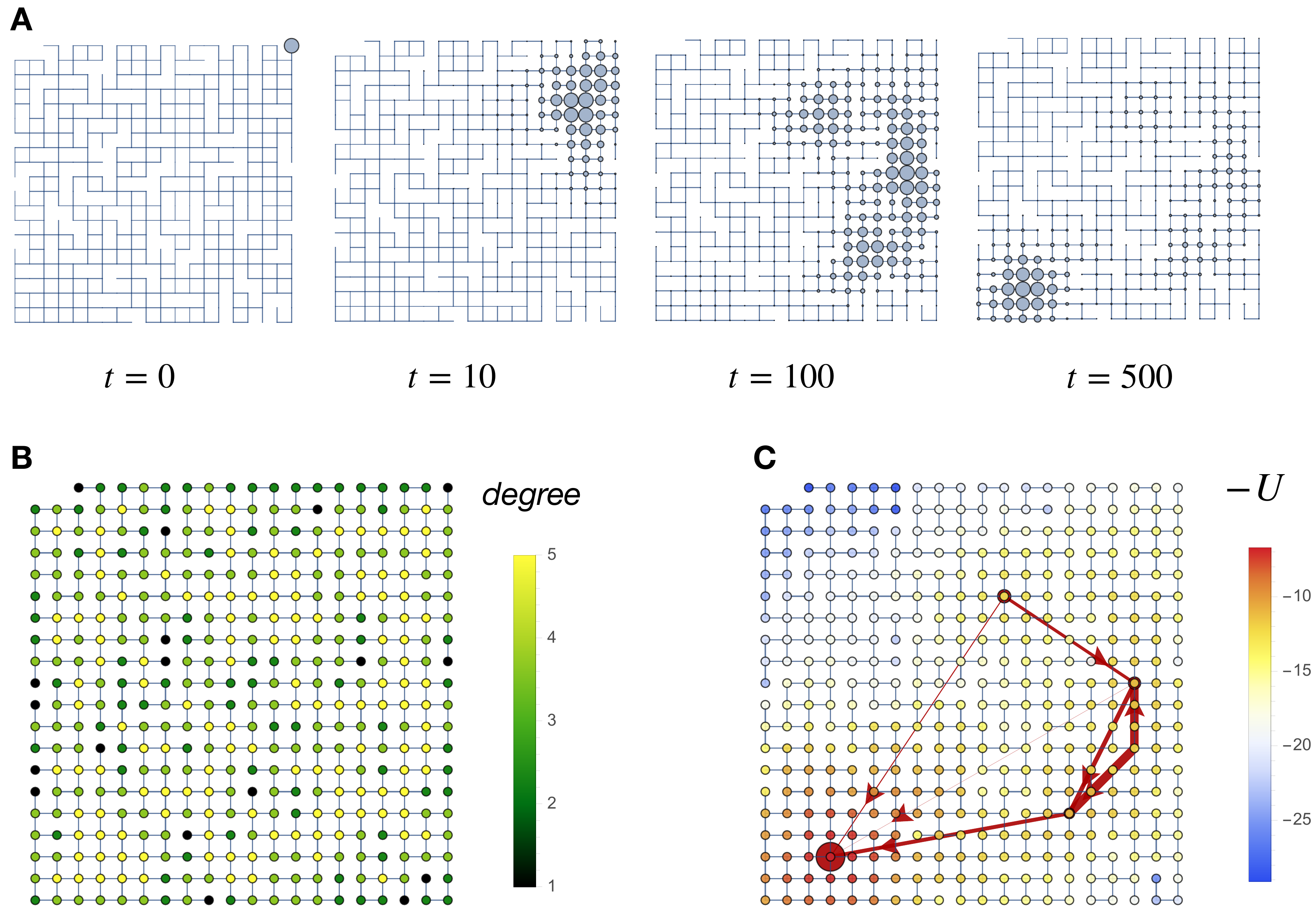}
\caption{{\emph{The hidden dynamics of graphs.} (A) Evolution of a localized
initial condition according to the equation~\(\dot{x}=Ax\)
where~\(A\) is the adjacency matrix of a graph (possibly up
to an additive scalar matrix), here a two-dimensional lattice
with~\(20\%\) of its edges removed. This process models e.g.
the early stages of an epidemic spread along a mixing
network~\protect\cite{Keeling_2005} or quasispecies evolution along a neutral
network~\protect\cite{van_Nimwegen_1999}. It is also directly related to the ``maximal
entropy random walk''~\protect\cite{Burda_2009}. The path taken by the
normalized solution (dots) exhibits intermittency, with long transients
in localized regions followed by fast jumps to remote regions. (B)
Predicting this dynamics is difficult from the map of local degrees,
which show no clear pattern or directionality. (C) Plotting instead the
effective potential~\(U\) and its basin hopping graph
(local minimal of~\(U\) and adjacency relations between
their basins of attraction, weighted by the height of barrier between
them, in red) reveals the location islands within the network and their
temporal ordering.~Trajectories can then be predicted by eye.~
{\label{121249}}%
}}
\end{center}
\end{figure}

Perhaps the simplest example of a disordered linear system is 
\begin{equation}\label{graphEq}
\dot{x}=Ax 
\end{equation}
with $A$ the adjacency matrix of an irregular graph such as the one shown in Fig. \ref{121249} and $x$ a vector of node weights. This equation describes e.g. the spread of a disease outbreak in a large susceptible population distributed over a mixing network: $x_i$ is number of susceptibles at site $i$, and we have $\dot{x}_i=(\alpha - \gamma) x_i+\beta \sum_{j\sim i} x_j$ where $\alpha$ is in the infection rate, $\gamma$ the recovery rate, $\beta$ the transmission rate along the edges of the network, and $j$ the neighbouring sites; up to an exponential factor this equation is equivalent to \eqref{graphEq}. As we will recall below, \eqref{graphEq} also arises naturally in the context of micro-evolution, where it describes molecular evolution along neutral networks \cite{Huynen_1996,van_Nimwegen_1999}.

Whether the application is epidemiological, evolutionary or else, the main problem associated to a problem such as  \eqref{graphEq} is to \textit{forecast} its evolution. Will its solutions localize? Where are its localization islands? Where and when will it jump? Although it completely specifies the system, the map of node degrees (Fig. \ref{121249}B) by itself provides little insight into the behavior of the solution $x(t)$, and these questions do not have a straightforward answer. Besides exact solutions in the simplest cases, there are few general principles to address these questions. An important one is Anderson's realization that localization is strong in low spatial dimensions \cite{Anderson_1958}, and, symmetrically, Eigen's prediction that high-dimensional hypercube graphs requires large fitness differences (strong disorder) to preserve structure in genotype populations \cite{Eigen_1971}.

The purpose of this paper is to address the forecasting problem for any autonomous linear system which preserves the positivity of its solutions. This includes the graph problem above, but also the parabolic Anderson model of branching random walks in random environments, introduced in  \cite{Zel_dovich_1987} and studied extensively in the mathematical literature as the prototype of intermittency \cite{knig2016}, or Eigen and Schuster's quasispecies model of molecular evolution \cite{schuster1989} (or any infinite-population mutation-selection model, e.g. Crow and Kimura's \cite{kimura2009}). Systems with time-dependent disorder, by contrast, are beyond the scope of the present approach. 

The problem of locating localization islands within a static disordered landscape has been addressed recently by Filoche and Mayboroda in the context of wave localization \cite{Filoche_2012}. The authors showed that the eigenvectors and eigenvalues of a disordered wave operator can be accurately estimated through a single function called the ``localization landscape''. The present work started as an attempt to apply the Filoche-Mayboroda localization landscape theory to intermittent dynamics, which formally resemble a Schr\"odinger-like equation in imaginary time. However, the localization landscape proves to be a poor predictor of transient localization and intermittency (see Fig. \ref{969087}C). One reason is that, unlike the quantum case, eigenvectors of the evolution operator do \textit{not} correspond to localized states; for starters, none of these eigenvectors (except one) is positive, and so they are not even admissible states.

%\footnote{It is possible to characterize localization islands in terms of local spectra defined by Dirichlet boundary conditions inboxes of growing sizes [Grtner2007]. The present approach is more straightforward, as it involves just one dominant eigenvector instead of a family of dominant eigenvectors with time-dependent boundary conditions.} 

Fortunately, there is a different notion of effective landscape which does predict intermittent dynamics, including localization islands and dominant transitions between them, accurately. The key to deriving it is a simple idea: intermittent dynamics lies in the conflicting effects of growth and transport (one induces localization, the other opposes it)---what if we could eliminate growth completely? I show that this is indeed always possible: every irreducible positive linear system can be mapped onto a Markov process with the same state space and modified transition rates. From this mapping several consequences follow immediately. First, intermittency in disordered media is seen to be the same phenomenon as metastability in multi-well potentials, as in reaction rate theory or biopolymer folding kinetics. Second, we can reduce (coarse-grain) the dynamics of intermittent systems along the metastable states of their associated Markov process, and estimate transition times using Arrhenius-Kramers formulas \cite{H_nggi_1990}. Third, we can construct an entropic Lyapunov function which decreases monotonically in any disordered landscape.

Interestingly, the mapping from linear systems to Markov process generalizes the construction of the so-called ``maximal entropy random walk'' (MERW), to which it reduces in the graph example of Fig. \ref{121249}. The MERW is defined as the stationary stochastic process on a graph which maximizes the entropy of its \textit{paths} \cite{Burda_2009}; it is therefore a canonical object on par with the simple random walk (which maximizes the entropy of individual \textit{jumps}). This connection provides a novel and intriguing interpretation to neutral evolution: it not mutations which are selected uniformly at random during neutral evolution, but rather mutational paths \cite{Smerlak_2020}. We will come back on this point in the last section.

\section{From positive linear systems to Markov processes}

\subsection{Main result}

An $n$-dimensional positive linear system (PLS) $\mathcal{M}$ is a linear dynamical system in the space of non-negative vectors, \textit{viz.} one that preserves the orthant $\mathbb{R}^n_+=\{x\in \mathbb{R}^n, x_1\geq 0, x_2\geq 0, \cdots,  x_n\geq0\}$. Such a system can be written as $\dot{x}=xM$ for some quasipositive (or Metzler) evolution matrix $M$, \textit{i.e.} one that satisfies $M_{ij}\geq 0$ for all $i\neq j$ \cite{Farina_2000,B_tkai_2017}. We assume this matrix to be irreducible, meaning that the associated directed graph is strongly connected. When $M$ is symmetric we write equivalently $\dot{x}=Mx$.\footnote{The theory generalizes to compact operators in Banach spaces via the Krein-Rutman theorem, although we do not pursue this here.} 

We may think of a PLS as a system coupling transport to growth: 
writing $g_i = \sum_j M_{ij}$ and $G = \textrm{diag}(g_i)_i$, we can decompose $M = G + T$ with $T$ the transition-rate matrix of a Markov process, \textit{viz}. $\sum_j T_{ij}=0$. In this representation, $T$ can be thought of as the transport component of the dynamics and $G$ as its growth component. In many cases of physical and biological interest, the transport component is simple, or even solvable; this is the case e.g. when $T$ is the  Laplacian matrix of a graph (a lattice or a Hamming graph, respectively), generating a simple random walk. Solving the (diagonal) growth component is of course also straightforward. The challenge of intermittency stems from the interaction between these two components. 

By Frobenius theory, $M$ has a dominant real eigenvalue $\Lambda$ such that all other eigenvalues $\lambda$ have $\Re  \lambda \leq \Lambda$, and the left and right eigenvectors associated to $\Lambda$, $L$ and $R$ respectively, are both positive  \cite{Farina_2000,B_tkai_2017}. (All other eigenvectors have nodes.) The left eigenvector $L$ describes the asymptotic state $\xi$ of the system,  \textit{viz.} $x(t)\sim e^{\Lambda t}L\equiv \xi(t)$ as $t\to\infty$. 
The basic observation of this paper is that any ($n$-dimensional, irreducible) PLS can be mapped onto an ($n$-state, irreducible) Markov process $\mathcal{Q}$ whose transition rate from $i$ to $j$ is given by 
\begin{equation}\label{toMarkov}
Q_{ij}=R_i^{-1}(M_{ij}-\Lambda \delta_{ij})R_j.
\end{equation}
The proof is trivial: given $i\neq j$,$M_{ij}\geq 0$ and $R_i>0$ imply $Q_{ij}\geq 0$; moreover we have by construction $\sum_j Q_{ij}=R_i^{-1}[(AR)_i - \Lambda R_i]=R_i^{-1}[\Lambda R_i - \Lambda R_i]=0$. We also easily check that if $M$ is itself the transition matrix of a Markov process (hence $\Lambda=0$ and $R=1$), then $T=M$, \textit{i.e.} \eqref{toMarkov} leaves Markov processes invariant. Finally, the invariant distribution of $\mathcal{Q}$ is $\pi_i=L_iR_i$ (choosing $L$ and $R$ such that $L\cdot R = \sum_i \pi_i = 1$).

One way to visualize the dynamics of $\mathcal{Q}$ is via the effective potential $U_i = -2 \log R_i$, which acts a biasing force with respect to the original transport process:
\begin{equation}
Q_{ij}=T_{ij}e^{(U_j - U_i)/2}. 
\end{equation}
In the special case where $T$ is symmetric (hence $\mathcal{Q}$ is reversible), the equilibrium distribution satisfying the detailed balance condition $\pi_iQ_{ij} = \pi_j Q_{ji}$ is the Gibbs distribution $\pi_i=e^{-U_i}$. In particular, when $T$ generates a simple random walk on a graph (\textit{i.e.} $T$ is the Laplacian of that graph), we can picture $\mathcal{Q}$ as a random walk in the potential $U_i$. Note that, from the perspective of the growth coefficients $\gamma_i$, this effective potential $U_i$ is non-local, \textit{i.e.} depends on all coefficients $(\gamma_i)_i$.

What makes the transformation \eqref{toMarkov} compelling from a dynamical perspective is that it preserves the time scales in the problem. Suppose the system is initialized in state $i$ at time $t=0$, and definite the local time $\tau_i$ as the time before $x_i(t)$ reaches its asymptotic value, \textit{i.e.} $\tau_i = \inf\,\{\tau>0, x_i(\tau)/\xi_i(\tau) = 1- \epsilon\}$. Since $x_i(t)$ is related to the density $p_i(t)$ of the Markov process $\mathcal{Q}$ through $x_i(t)=e^{\Lambda t}R_i^{-1}p_i(t)$, we see that $\tau_i$ is also the local time under $\mathcal{Q}$: $\tau_i = \inf\,\{\tau>0, p_i(\tau)/\pi_i = 1- \epsilon\}$. As a result, the relative stability of a state under $\mathcal{M}$ can be assessed by measuring its relative stability under $\mathcal{Q}$.

\subsection{Related constructions}

While it is new as a general tool to forecast evolution, the  transformation \eqref{toMarkov} has multiple precedents in the literature. $(i)$ When $M$ is the adjacency matrix of a simple graph $G$, \eqref{toMarkov} defines the (continuous-time) ``maximal entropy random walk'' on $G$ \cite{Ruelle_2004,Burda_2009}. This is the case of the epidemiological example given in the introduction. The MERW has found applications in networks science, e.g. for link prediction \cite{Li_2011} or community detection \cite{Ochab_2013}; it is also directly related to neutral evolution \cite{Smerlak_2020}, as explained below. $(ii)$ In demography and ecology, the dynamics of age-structured populations can be described using a PLS known as the Leslie model. Ref. \cite{Tuljapurkar_1982} used equation \eqref{toMarkov} to relate this model to a Markov chain on the set of age classes, which helped clarified the meaning of Demetrius' notion of ``population entropy'' \cite{Demetrius_1974}; this was later extended to more general matrix population models in \cite{Bienvenu_2017}. $(iii)$ Within the stochastic approach to quantum mechanics, the Markov process defined by \eqref{toMarkov} for $M=H$, the Hamiltonian of a non-relativistic particle, was used by Yasue to study tunneling rates across a potential barrier \cite{Yasue_1978}. $(iv)$ Similarly, Fokker-Planck equations are often studied in terms of Hamiltonian operators and their spectra through a transformation which is essentially inverse to \eqref{toMarkov}, see e.g. \cite{Risken_1989}. $(v)$ In the theory of classical  \cite{chetrite2015nonequilibrium} and quantum \cite{macieszczak2020theory} Markov processes, the transformation \eqref{toMarkov} arises upon conditioning as a `generalized Doob transform' \cite{doob1957conditional}. $(vi)$ In the context of evolutionary dynamics, the time-reversal of the process \eqref{toMarkov}, given by $Q'_{ij}=L_i^{-1}(M_{ji}-\Lambda \delta_{ij})L_j$ (and coinciding with \eqref{toMarkov} when $M$ is symmetric), was used in \cite{Hermisson_2002} to characterize the ancestral distribution at mutation-selection equilibrium. $(vii)$ Also in an evolutionary context, McCandlish has suggested visualizing fitness landscapes in terms of a Markov chain in genotype space \cite{McCandlish_2011}. While his approach is both mathematically and biologically distinct (in particular, McCandlish focuses on monomorphic populations in the weak mutation regime, instead of quasispecies-type dynamics below), the motivation is the same as here: turning the raw growth data in $M$ into a dynamically meaningful transport process. 
\begin{figure*}[t!]
\begin{center}
\includegraphics[width=0.98\textwidth]{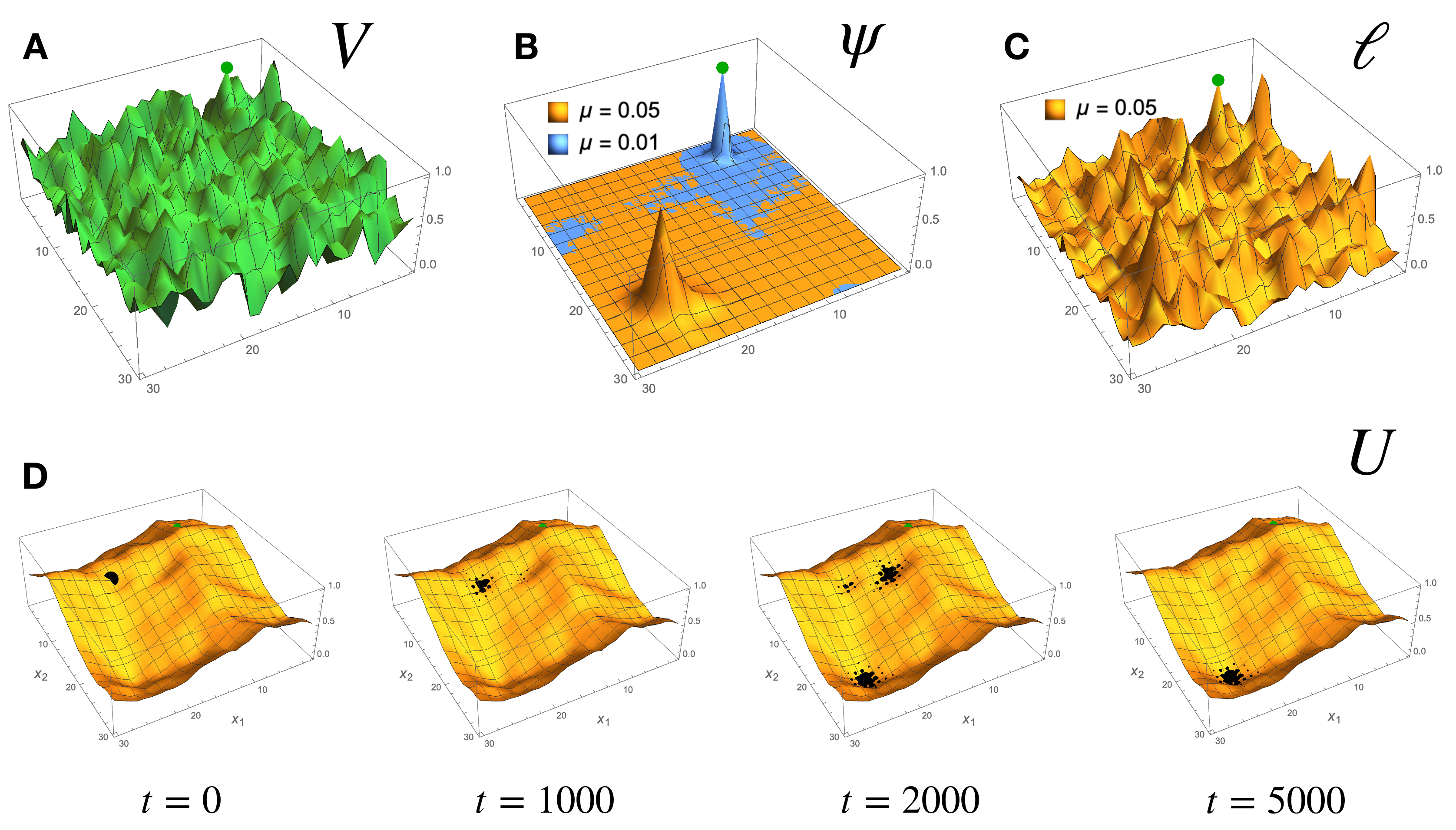}
\caption{{\emph{Smoothness within ruggedness.}~(A) A Gaussian potential landscape
over a 2d lattice with periodic boundary conditions, the typical setup
for the parabolic Anderson model. The global potential minimum is
indicated by a green dot. (B) The ground state~\(\psi\) for
that landscape at two values of the diffusivity~\(\mu\),
exhibiting localization at the global peak (\(\mu=0.01\)) or
somewhere else (\(\mu=0.05\)). (C) Although useful in predicting
the location of the dominant eigenvectors of the Anderson
Hamiltonian~\(H\), the Filoche-Mayboroda localization
landscape~\(\ell\) does not help much to predict the
intermittent dynamics dictated by~\(\dot{x}=Hx\). (D) By contrast,
the effective potential~\(U\) has a smooth structure with
clearly defined local minima, which the solution (black dots) follows
closely. Note how the system conspicuously moves~\emph{away} from the
global peak.~
{\label{969087}}%
}}
\end{center}
\end{figure*}

\section{Intermittency as metastability}

Upon the transformation of the original PLS into a Markov chain, intermittency naturally reduces to the well-known concept of metastability, studied since van't Hoff, Arrhenius and Kramers. This has several immediate advantages.

\subsection{Locating localization islands}

The problem of reducing a metastable Markov process to a simpler, coarser process is a classic problem in non-equilibrium statistial mechanics. The best understood case corresponds to reversible Markov processes, particularly random walks in external potentials $U_i$. We will focus on this case here, noting that the more general theory remains an active field of mathematical research \cite{Landim_2019}. 

The key principle of Markovian metastability is that the systems favors regions of low potential, with transitions between local minima (and their basins of attraction) exponentially suppressed in the barrier height $\Delta U = U_s - U_m$, where $U_m$ is the value at the minimum and $U_s$ the value at the lowest saddle above that minimum; this is the content of Kramers' law \cite{H_nggi_1990}. This leads us to an important insight: the localization islands of a symmetric PLS are centered on those local minima of the effective potential $U$ which have $\Delta U\gg1$; the time scale between transitions is $e^{-\Delta U}$. Figs. \ref{121249}-\ref{969087} confirm this prediction in cases where the disorder comes from the underlying graph or some potential defined on that graph, respectively.  

\subsection{Coarse-grained trajectories}

Perhaps the most important implication of this identification of localization islands as minima of the effective potential is that it allows to reduce, or coarse-grain, the dynamics of the system. Coarse-graining is the ultimate goal of complex systems theory, for it shows which features of a complex system are relevant for its emergent behavior, and which are not;---it helps see the order hidden within disorder. Here, this problem is immediately solved: all states $i$ that belong to the basin of attraction of a potential minimum can be collapsed into a single macrostate; adjacent basins separated by a low barrier $\Delta U<1$ can further be collapsed, as transitions between them are frequent and time scales are not decoupled. Finally,  macrostates can be connected according to the lowest of the high barriers that separate them, following the large deviation principle for Markov processes to which transitions between basins are exponentionally (in $\Delta U$) more likely to happen along the lowest saddles. 

This leads to a representation of the system in terms of a ``basin hopping graph'' (BHG), a concept introduced in the concept of RNA folding kinetics \cite{Kuchar_k_2014}. One such BHG is shown in Fig. \ref{121249}. It is clear from that figure that, while inferring the dynamics from the original graph is difficult, it is straightforward from the BHG. Moreover, unlike other representations of the effective potential $U$, BHGs are suited to high-dimensional problems, e.g. evolutionary ones (where the underlying graph is typically a hypercube graph). 

\subsection{An entropic Lyapunov function}

Thirdly, the Markovian interpretation immediately provides a general Lyapunov function for positive linear systems. Finding such a Lyapunov function has been a focus of special interest in the context of evolutionary dynamics, perhaps because of the impact (but also the limitation) of Fisher's fundamental theorem of natural selection \cite{Fisher_1930}; see e.g. \cite{Sella_2005} in the weak selection weak mutation regime or \cite{Jones_1978} in the context of quasi-species theory. 

It is a general property of irreducible Markov process that the relative entropy between the probability density and its equilibrium $D[p(t)\Vert \pi]$ decreases monotonically. Translating this into the original variable $x(t)$, we see immediately that $F(t)=D[e^{-\Lambda t}Rx(t)\Vert LR]$ is a Lyapunov function for the system $\mathcal{M}$ (Fig. \ref{962302}B-C).

\section{Examples}

\subsection{Parabolic Anderson model}

The parabolic Anderson model (PAM) starts from a random Schr\"odinger operator $H = \kappa \Delta + V$, where $\Delta$ is the Laplacian on a $d$-dimensional lattice and $V$ a random field (here we take $d=2$ and $V$ a Gaussian process), and considers the corresponding positive linear system 
\begin{equation}\label{PAM}
\dot{x}=Hx = \mu \Delta x + V x
\end{equation}
with $\mu>0$ a diffusion coefficient. This equation was initially motivated by certain problems in chemical and nuclear kinetics and in magnetohydrodynamics \cite{Zel_dovich_1987}, and is now extensively studied as a prototype of intermittent behavior \cite{G_rtner}. It is also related, via a Wick rotation, to Anderson's model of electron localization \cite{Anderson_1958}. Finally, the PAM (or any other imaginary-time Schr\"odinger equation) can be interpreted in terms of branching random walks; this interpretation underlies  diffusion Monte Carlo techniques in quantum many-body theory \cite{Reynolds_1990}. 

Assume given the ground state $\psi$ of $H$. (In SI we recall an analytical approach to computing $\psi$ commonly used in condensed-matter physics, the forward scattering approximation. Alternatively, $\psi$ can be computed using exact numerical diagonalization or via quantum Monte Carlo techniques.) Then, applying the transformation \eqref{toMarkov}, we see that \eqref{PAM} is equivalent to a random walk in the effective potential $U=-2 \log \psi$. Thanks to the smoothing effect of the Laplacian, this potential is much more regular than $V$, with a lot fewer  minima. This makes the dynamics of the PAM much more straightforward to analyze: it suffices to locate these minima and their basins of attraction to predict the direction (and times, through Kramers' formula) of intermittent transitions. 

Fig. \ref{969087} displays an example of this dramatic simplification in the dynamics of the PAM. Fig. \ref{969087}A shows a disorderd potential $V$, here sampled from a Gaussian process with unit correlation length on a $2d$ lattice with periodic boundary conditions. Predicting the evolution of the distribution of mass $x(t)$ from the localized initial condition $x(0)$ (green dot) is clearly a difficult problem, at least based on visual inspection of the potential; moreover Fig. \ref{969087}B shows that the answer depends sensitively on the diffusivity $\mu$. But in Fig. \ref{969087}C, where the effective potential is represented instead of $V$, this problem becomes trivial: the mass will flow to the global minimum of $U$ after spending time in a neighboring local minimum. 

It is interesting to constrast this picture with the one obtained using the Filoche-Mayboroda localization landscape $\ell$, defined by the linear equation $H\ell = -1$ and capturing the low-lying eigenstates of $H$ near its peaks \cite{Filoche_2012}. Since computing the solution of linear system is usually faster than computing an eigenvector, the Filoche-Mayboroda approach could in principle be even more powerful to understand the dynamics of the PAM than the present one relying on $U$. Yet, Fig. \ref{969087} shows that the evolution of the PAM is less transparent viewed through  $\ell$ than it is through $U$, in part because $U$ is smoother than $\ell$. 

\begin{figure}[t!]
\begin{center}
\includegraphics[width=0.98\columnwidth]{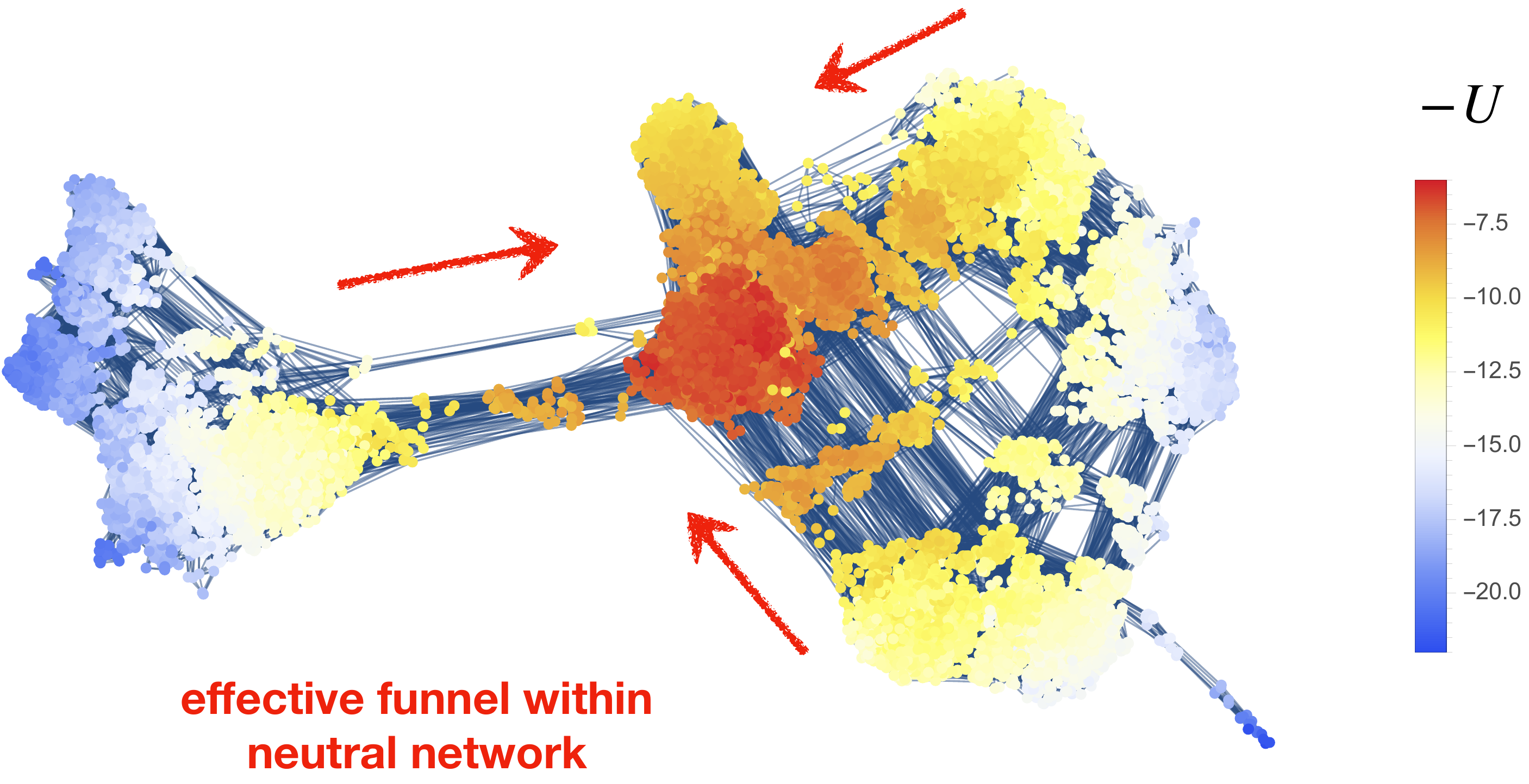}
\caption{{\emph{Funnels within neutral networks.~}The neutral network of a short
RNA secondary structure~\protect\cite{van_Nimwegen_1999}, with each node representing
a distinct sequence folding into that structure, over~\(10^4\)
in tihs case. Neutral evolution is usually pictured as the blind
exploration of neutral networks. Coloring the network with the effective
potential shows that this picture is wrong, at least when populations
are large and mutational robustness becomes an evolutionary advantage:
populations evolving in this network are attracted towards a small core
of mutationally robust sequences along a funnel-shaped structure hidden
within the network.~
{\label{117230}}%
}}
\end{center}
\end{figure}

\subsection{Neutral evolution}

Quasispecies theory studies the evolution of large populations of error-prone replicators such as auto-catalytic molecules \cite{schuster1989} or viruses \cite{Domingo_2019}. Each genotype $i$ has Malthusian fitness (exponential growth rate) $\phi_i$ and mutations with a transition rate matrix which may be approximated by $\mu\Delta$ where $\Delta$ is the Laplacian matrix of a graph $G$ with genotypes as nodes and point mutations as edges. (For binary sequences this graph is a hypercube.) The quasispecies equation is non-linear, but a simple transformation maps it to the PLS $\dot{x}=\mu x\Phi \Delta$ with $\Phi = \textrm{diag}(\phi_i)_i$ a diagonal matrix of fitness values. (An alternative model of mutation-selection dynamics, the Crow-Kimura model \cite{kimura1970}, gives a symmetric evolution matrix $M = \Phi + \mu\Delta$, \textit{i.e.} one that is formally identical to the PAM on a Hamming graph \cite{Avena_2020}. In this sense, mutation-selection dynamics is more similar to \textit{many-body} localization \cite{Abanin_2019} than to Anderson localization \cite{Anderson_1958}.)

It is generally accepted that much of micro-evolution is in fact neutral \cite{Nei_2005}: mutations either leave the fitness of the replicator unchanged, or they are so strongly deleterious that they leave no progeny \cite{Eyre_Walker_2007}. (The corresponding fitness landscapes are sometimes called `holey' \cite{Gavrilets_1997}.) This situation corresponds to a fitness matrix $\Phi$ which takes two values: $0$ for viable genotypes and $-\infty$ for non-viable ones. Inserting this into the quasispecies equations amounts to projecting it to the subgraph $G_0$ of viable genotypes (or neutral network). Denoting $A_0$ its degree matrix, this results in $\dot{x} = \mu A_0 x$, \textit{i.e.} neutral evolution is completely determined by the topology of the corresponding neutral network \cite{van_Nimwegen_1999}. 

This observation shows that neutral evolution is formally the same process as the epidemiological example discussed above; in particular, the Markov process associated to it is the MERW on $G_0$. One consequence is that the existence of a percolating subgraph of viable genotypes within genotype space is necessary \cite{MAYNARD_SMITH_1970}, but not sufficient \cite{Shorten_2017, Smerlak_2020}, for open-ended evolution: just like an epidemic outbreak can attack localized hotspots rather than spread uniformly along a mixing network, a large, neutrally evolving population can fail to explore its genotype space due to MERW localization \cite{Burda_2009}.  

The second consequence of the present Markovian approach is that we may describe the (slow) neutral evolution of molecules in the same way as its (fast) folding kinetics. Folding kinetics involves a random walk in conformational space biased by potential energy; similarly, neutral evolution can be viewed as a random walk in genotype space biased by the effective potential $U=-2\log E$, where $E$ is the mutation-selection equilibrium distribution, \textit{i.e.} the vector of eigenvector centralities for the graph $G_0$. 

Fig. shows the neutral network of the RNA secondary structured studied in \cite{van_Nimwegen_1999,McCandlish_2011} graded by the effective potential $U$, showing that neutral evolution of this RNA structure is directed towards a highly connected core \cite{Shorten_2017}. The ``evolution of mutational robustness'' was predicted in \cite{van_Nimwegen_1999} from the properties of the asympotic mutation-selection balance $E$. The present approach reveals how such robustness evolves on the way to  equilibrium, as the endpoint of a funnel structure hidden within the neutral network.

\subsection{NKp landscapes}

\begin{figure*}[t!]
\begin{center}
\includegraphics[width=0.98\textwidth]{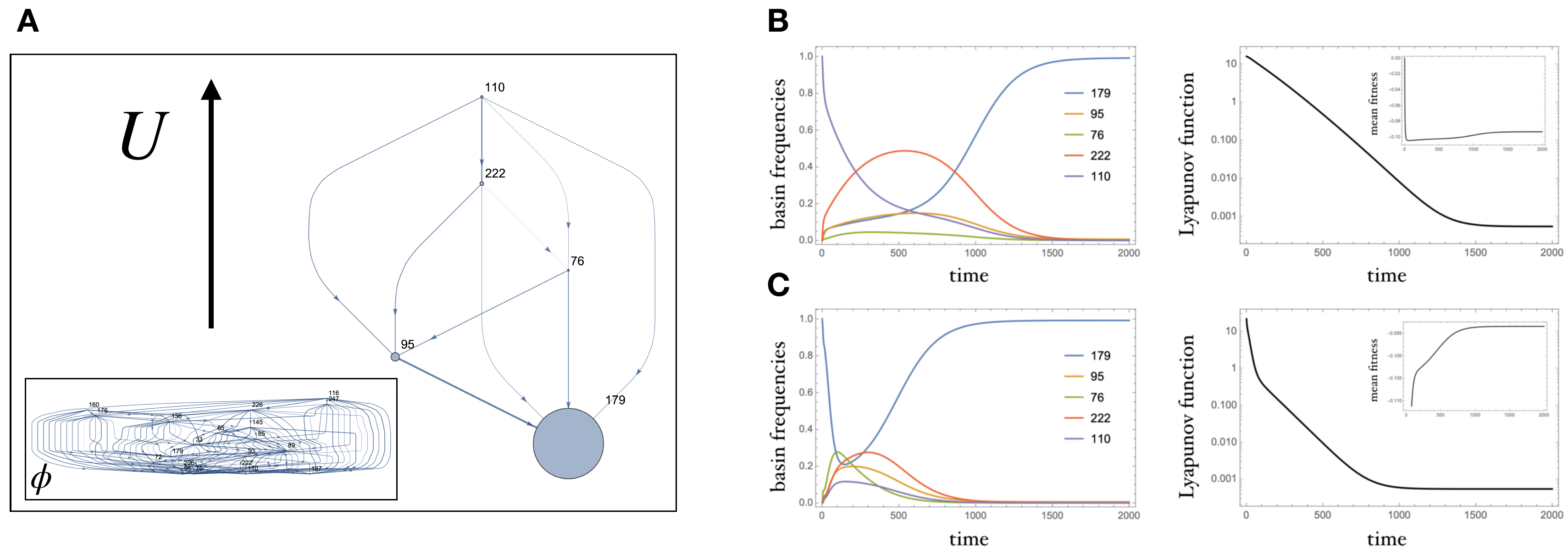}
\caption{{\emph{Predicting evolution in a rugged landscape with neutrality:}
the~\emph{NKp} genotypic landscape with~\(2^{8}=256\) types. A: The
basin hopping graph (BHG) for the effective potential (here for a
genomic mutation rate~\(\mu=0.1\)), in which nodes are basins of
attractions of potential minima and edges adjacency relations between
basins weighted by the barrier height. This coarse-grained
representation is much simpler than the BHG for the fitness landscape,
which has~\(20\) local maxima (inset)---and immediately
predictive. B: A sample coarse-grained evolutionary trajectory, obtained
by integration of the mutation-selection equation. A population
initially concentrated in basin 110 moves towards basin 179 through
basins 222 and 95, as suggested by the BHG on the left. This happens in
spite of the fact that 110 is a global fitness maximum and mean fitness
decreases in time. C: A population initially concentrated in basin 179
spreads to neighbouring basins before concentrating again; in spite of
this behavior the Lyapunov function~\(F\left(t\right)\) decreases
monotonically over time.~
{\label{962302}}%
}}
\end{center}
\end{figure*}

Finally, let us consider a synthetic fitness landscape with tunable neutrality and ruggedness, the $NKp$ landscape \cite{Kauffman_1987,barnett1998}. Here, the fitness $\phi_i$ of a binary sequency $i$ of length $N$ is given as the sum of $K$ random fitness correlations, correlated along the edges of a cyclic graph; with probability $p$, these contributions are zeroed out, leading to extended neutral networks within the landscape. One instance of such landscapes with $N=8$, parameter $K=6$ and $p=0.7$ is showed in Fig. \ref{962302}. This instance has $20$ local maxima and an error threshold at $\mu_c\simeq 0.2$. Comparing the basin hopping graphs of the fitness landscape $\phi$ and of the potential landscape $U$ reveals that most of the complexity of the former is spurious. Moreover, coarse-grained evolutionary trajectories, described by the coarse-grained frequencies, are consistent with the succession of transitions predicated by the basin hopping graph of $U$: a population initially concentrated around the genotype $110$ (a global fitness maximum) will evolve towards the flatter genotype $179$ via the basins of $222$ and $95$. One also checks that the Lyapunov function $G(t)$ decreases monotonically also when the population mean fitness does not (Fig. \ref{962302}B) and when the basin frequencies have strongly non-monotonic behavior (Fig. \ref{962302}C).

\section{Discussion}
Intermittency is a common phenomenon in disordered systems powered by growth and transport processes, be them physical, chemical, biological or socio-economical. In this paper I have showed that intermittency is fundamentally the same phenomenon as Markovian metastability: a viral population which suddenly hops to a new peak after crossing a valley in its fitness landscape is not different than a Brownian particle which suddenly hops over a potential barrier through thermal fluctuations; in the language of optimization theory, a ``Darwin'' search strategy (based on competitive selection) is not different than a ``Boltzmann'' strategy (based on energy minimization with thermal noise) \cite{Dunkel_2003}. What is more, we have seen that the early progression of an epidemic on a mixing network, neutral evolution in a holey fitness or the entropy-rate maximizing MERW are all the same basic process, canonically associated to a graph through its adjacency matrix. The Markovian approach thus provides a unifying perspective on various concepts and methods across science. This can lead to new insights: to my knowledge, the intermittency of epidemic spreading on disordered low-dimensional lattices, with the outbreak jumping from hotspot to hotspot with an infinite susceptible population and without any intervention, has not be noted before. 

But the value of the Markovian approach to intermittency is not merely conceptual---it also greatly reduces the complexity of the problem, combinatorially, dynamically and computationally. First, visualizing large, high-dimensional landscapes (such as molecular fitness landscapes) is by nature difficult. Using standards tools designed for folding kinetics such as the barrier tree package \cite{Flamm_2002}, we can efficiently represent such a landscape as a small graphs with metastable basins---rather than individual states---as nodes. Second, even when visualizing the disordered landscape is straightforward, as in the maze of Fig. \ref{121249} or the Gaussian potential of Fig. \ref{969087}, it is often challenging to predict the motion of the system, which does not optimize any local quantity (degree or potential). Trading the landscape for the effective potential (and its reduced representations, barrier tree or BHG) amounts to cutting through the noise and  reveals the hidden pathways within the system. Third, solving exactly the evolutionary equation $\dot{x}=xM$ involves computing a matrix exponential for each time $t$ and each initial condition $x(0)$. Matrix exponentiation is a costly operation, with computational complexity comparable to that of the complete diagonalization of $M$ \cite{Moler_2003}; by contrast, computing the effective potential amounts to obtaining just the dominant eigenvector of a (usually sparse) matrix, which is much faster thanks to iterative methods such as Arnoldi's or Lanczos' algorithms. Coarse-graining that potential with a flooding algorithm is faster still  \cite{Flamm_2002}. 

As already noted, a major area of application of this work concerns evolutionary dynamics in complex fitness landscapes \cite{Smerlak_2019}. While much progress has been made in recent years thanks to improved sequencing technologies, empirical data on real fitness landscapes remains scarse \cite{Szendro_2013}. It is not inconceivable, however, that this situation changes in the near future, e.g. when the local fitness landscape of a viral or cancer genome is measured comprehensively. When this time comes, the bottleneck to evolutionary prediction \cite{L_ssig_2017} might become the computation of likely evolutionary trajectories from fitness data. The present theory provides a method to do so efficiently.

\section*{Acknowledgements}

I thank D. McCandlish and J. P. Garrahan for pointing me to Ref. \cite{Hermisson_2002,McCandlish_2011,Bienvenu_2017} and Ref. \cite{chetrite2015nonequilibrium, macieszczak2020theory} respectively, and M. Kenmoe, A. Klimek, M. L{\"a}ssig, O. Rivoire, D. Saakian and A. Zadorin for useful discussions. Funding for this work was provided by the Alexander von Humboldt Foundation in the framework of the Sofja Kovalevskaja Award endowed by the German Federal Ministry of Education and Research.

\bibliographystyle{pnas-new}
\bibliography{biblio.bib}

\begin{thebibliography}{10}

\bibitem{eldredge1977}
Gould SJ, Eldredge N (1977) {Punctuated Equilibria: The Tempo and Mode of
  Evolution Reconsidered}.
\newblock {\em Paleobiology} 3:115--151.

\bibitem{chao1999}
Burch CL, Chao L (1999) {Evolution by small steps and rugged landscapes in the
  RNA virus phi6}.
\newblock {\em Genetics} 151:921--927.

\bibitem{Lenski_1994}
Lenski RE, Travisano M (1994) {Dynamics of adaptation and diversification: a
  10,000-generation experiment with bacterial populations.}
\newblock {\em Proc. Natl. Ac. Sci. USA} 91(15):6808--6814.

\bibitem{Lessler_2017}
Lessler J, Azman AS, McKay HS, Moore SM (2017) {What is a Hotspot Anyway?}
\newblock {\em Am. J. Trop. Med. Hyg.} 96(6):1270--1273.

\bibitem{Zanette_1997}
Zanette DH, Manrubia SC (1997) {Role of Intermittency in Urban Development: A
  Model of Large-Scale City Formation}.
\newblock {\em Phys. Rev. Lett.} 79(3):523--526.

\bibitem{Shandarin_1989}
Shandarin SF, Zeldovich YB (1989) {The large-scale structure of the universe:
  Turbulence, intermittency, structures in a self-gravitating medium}.
\newblock {\em Rev. Mod. Phys.} 61(2):185--220.

\bibitem{Zel_dovich_1987}
Zeldovich YB, Molchanov SA, Ruzma{\u{\i}}kin AA, Sokolov DD (2007)
  {Intermittency in random media}.
\newblock {\em Sov. Phys. Usp.} 30(5):353--369.

\bibitem{Huynen_1996}
Huynen MA, Stadler PF, Fontana W (1996) {Smoothness within ruggedness: the role
  of neutrality in adaptation.}
\newblock {\em Proc. Natl. Ac. Sci. USA} 93(1):397--401.

\bibitem{van_Nimwegen_1999}
Van~Nimwegen E, Crutchfield JP, Huynen M (1999) {Neutral evolution of
  mutational robustness}.
\newblock {\em Proc. Natl. Ac. Sci. USA} 96(17):9716--9720.

\bibitem{Anderson_1958}
Anderson PW (1958) {Absence of Diffusion in Certain Random Lattices}.
\newblock {\em Phys. Rev.} 109(5):1492--1505.

\bibitem{Eigen_1971}
Eigen M (1971) {Selforganization of matter and the evolution of biological
  macromolecules}.
\newblock {\em Naturwissenschaften} 58(10):465--523.

\bibitem{Keeling_2005}
Keeling MJ, Eames KTD (2005) {Networks and epidemic models}.
\newblock {\em J. R. Soc. Interface} 2(4):295--307.

\bibitem{Burda_2009}
Burda Z, Duda J, Luck JM, Waclaw B (2009) {Localization of the Maximal Entropy
  Random Walk}.
\newblock {\em Phys. Rev. Lett.} 102(16):549.

\bibitem{knig2016}
K{\"o}nig W (2016) {\em {The Parabolic Anderson Model}}.
\newblock (Springer).

\bibitem{schuster1989}
Eigen M, McCaskill J, Schuster P (1989) {The molecular quasi-species}.
\newblock {\em Adv. Chem. Phys.} 75:149--263.

\bibitem{kimura2009}
Crow JF, Kimura M (2009) {\em {An Introduction to Population Genetics Theory}}.
\newblock (The Blackburn Press).

\bibitem{Filoche_2012}
Filoche M, Mayboroda S (2012) {Universal mechanism for Anderson and weak
  localization}.
\newblock {\em Proc. Natl. Ac. Sci. USA} 109(37):14761--14766.

\bibitem{H_nggi_1990}
H{\"a}nggi P, Talkner P, Borkovec M (1990) {Reaction-rate theory: fifty years
  after Kramers}.
\newblock {\em Rev. Mod. Phys.} 62(2):251--341.

\bibitem{Smerlak_2020}
Smerlak M (2020) {Localization of neutral evolution: selection for mutational
  robustness and the maximal entropy random walk}.

\bibitem{Farina_2000}
Farina L, Rinaldi S (2000) {\em {Positive Linear Systems}}, Theory and
  Applications.
\newblock (John Wiley {\&} Sons, Inc., Hoboken, NJ, USA).

\bibitem{B_tkai_2017}
B{\'a}tkai A, Fijav{\v z} MK, Rhandi A (2017) {Positive Linear Systems} in {\em
  Positive Operator Semigroups}.
\newblock (Springer International Publishing, Cham), pp. 93--105.

\bibitem{Note1}
(year?).
\newblock The theory generalizes to compact operators in Banach spaces via the
  Krein-Rutman theorem, although we do not pursue this here.

\bibitem{Ruelle_2004}
Ruelle D (2010) {\em {Thermodynamic Formalism}}, The Mathematical Structure of
  Equilibrium Statistical Mechanics.
\newblock (Cambridge University Press) Vol.{}~19, 2 edition.

\bibitem{Li_2011}
Li RH, Yu JX, Liu J (2011) {Link prediction} in {\em CIKM '11}.
\newblock (ACM Press, New York, New York, USA), p. 1147.

\bibitem{Ochab_2013}
Ochab JK, Burda Z (2013) {Maximal entropy random walk in community detection}.
\newblock {\em Eur. Phys. J. Spec. Top.} 216(1):73--81.

\bibitem{Tuljapurkar_1982}
Tuljapurkar SD (1982) {Why use population entropy? It determines the rate of
  convergence}.
\newblock {\em J. Math. Biol.} 13(3):325--337.

\bibitem{Demetrius_1974}
Demetrius L (1974) {Demographic Parameters and Natural Selection}.
\newblock {\em Proc. Natl. Ac. Sci. USA} 71(12):4645--4647.

\bibitem{Bienvenu_2017}
Bienvenu F, Ak{\c c}ay E, Legendre S, McCandlish DM (2017) {The genealogical
  decomposition of a matrix population model with applications to the
  aggregation of stages}.
\newblock {\em Theor. Pop. Biol.} 115:69--80.

\bibitem{Yasue_1978}
Yasue K (1978) {Detailed Time-Dependent Description of Tunneling Phenomena
  Arising from Stochastic Quantization}.
\newblock {\em Phys. Rev. Lett.} 40(11):665--667.

\bibitem{Risken_1989}
Risken H (1989) {\em {The Fokker-Planck Equation}}, Springer Series in
  Synergetics.
\newblock (Springer Berlin Heidelberg, Berlin, Heidelberg) Vol.{}~18.

\bibitem{chetrite2015nonequilibrium}
Chetrite R, Touchette H (2015) Nonequilibrium markov processes conditioned on
  large deviations in {\em Annales Henri Poincar{\'e}}.
\newblock (Springer), Vol.{}~16, pp. 2005--2057.

\bibitem{macieszczak2020theory}
Macieszczak K, Rose DC, Lesanovsky I, Garrahan JP (2020) Theory of classical
  metastability in open quantum systems.
\newblock arXiv:2006.01227.

\bibitem{doob1957conditional}
Doob JL (1957) Conditional brownian motion and the boundary limits of harmonic
  functions.
\newblock {\em Bulletin de la Soci{\'e}t{\'e} Math{\'e}matique de France}
  85:431--458.

\bibitem{Hermisson_2002}
Hermisson J, Redner O, Wagner H, Baake E (2002) {Mutation{\textendash}Selection
  Balance: Ancestry, Load, and Maximum Principle}.
\newblock {\em Theor. Pop. Biol.} 62(1):9--46.

\bibitem{McCandlish_2011}
McCandlish DM (2011) {VISUALIZING FITNESS LANDSCAPES}.
\newblock {\em Evolution} 65(6):1544--1558.

\bibitem{Landim_2019}
Landim C (2019) {Metastable Markov chains}.
\newblock {\em Probab. Surveys} 16(0):143--227.

\bibitem{Kuchar_k_2014}
Kuchar{\'\i}k M, Hofacker IL, Stadler PF, Qin J (2014) {Basin Hopping Graph: a
  computational framework to characterize RNA folding landscapes}.
\newblock {\em Bioinformatics} 30(14):2009--2017.

\bibitem{Fisher_1930}
Fisher RA (year?) {\em {The genetical theory of natural selection.}}
\newblock (Clarendon Press, Oxford).

\bibitem{Sella_2005}
Sella G, Hirsh AE (2005) {The application of statistical physics to
  evolutionary biology}.
\newblock {\em Proc. Natl. Ac. Sci. USA} 102(27):9541--9546.

\bibitem{Jones_1978}
Jones BL (1978) {Some principles governing selection in self-reproducing
  macromolecular systems}.
\newblock {\em J. Math. Biol.} 6(2):169--175.

\bibitem{G_rtner}
G{\"a}rtner J, K{\"o}nig W (2005) {The Parabolic Anderson Model} in {\em
  Interacting Stochastic Systems}.
\newblock (Springer-Verlag, Berlin/Heidelberg), pp. 153--179.

\bibitem{Reynolds_1990}
Reynolds PJ, Tobochnik J, Gould H (1990) {Diffusion Quantum Monte Carlo}.
\newblock {\em Comput. Phys.} 4(6):662.

\bibitem{Domingo_2019}
Domingo E, Perales C (2019) {Viral quasispecies}.
\newblock {\em PLoS Genet} 15(10):e1008271.

\bibitem{kimura1970}
Crow JF, Kimura M (1970) {\em {An Introduction to Population Genetics Theory}}.
\newblock (Harper and Row, New York).

\bibitem{Avena_2020}
Avena L, G{\"u}n O, Hesse M (2020) {The parabolic Anderson model on the
  hypercube}.
\newblock {\em Stoch. Proc. Appl.} 130(6):3369--3393.

\bibitem{Abanin_2019}
Abanin DA, Altman E, Bloch I, Serbyn M (2019) {Colloquium: Many-body
  localization, thermalization, and entanglement}.
\newblock {\em Rev. Mod. Phys.} 91(2):168.

\bibitem{Nei_2005}
Nei M (2005) {Selectionism and Neutralism in Molecular Evolution}.
\newblock {\em Molecular Biology and Evolution} 22(12):2318--2342.

\bibitem{Eyre_Walker_2007}
Eyre-Walker A, Keightley PD (2007) {The distribution of fitness effects of new
  mutations}.
\newblock {\em Nat. Rev. Genet.} 8(8):610--618.

\bibitem{Gavrilets_1997}
Gavrilets S (1997) {Evolution and speciation on holey adaptive landscapes}.
\newblock {\em Trends Ecol. Evol.} 12(8):307--312.

\bibitem{MAYNARD_SMITH_1970}
Maynard~Smith J (1970) {Natural Selection and the Concept of a Protein Space}.
\newblock {\em Nature} 225(5232):563--564.

\bibitem{Shorten_2017}
Shorten D, Nitschke G (2017) {The Two Regimes of Neutral Evolution:
  Localization on Hubs and Delocalized Diffusion} in {\em Applications of
  Evolutionary Computation}.
\newblock (Springer International Publishing, Cham), pp. 310--325.

\bibitem{Kauffman_1987}
Kauffman S, Levin S (1987) {Towards a general theory of adaptive walks on
  rugged landscapes}.
\newblock {\em J. Theor. Biol.} 128(1):11--45.

\bibitem{barnett1998}
Barnett L (1998) {Ruggedness and neutrality - the NKp family of fitness
  landscapes} in {\em Adami, C., Belew, R.K., Kitano, H., Taylor, C. (eds.)
  Proc. Artificial Life VI}.
\newblock pp. 18--27.

\bibitem{Dunkel_2003}
Dunkel J, Ebeling W, Schimansky-Geier L, H{\"a}nggi P (2003) {Kramers problem
  in evolutionary strategies}.
\newblock {\em Phys. Rev. E} 67(6):284.

\bibitem{Flamm_2002}
Flamm C, Hofacker IL, Stadler PF, Wolfinger MT (2002) {Barrier Trees of
  Degenerate Landscapes}.
\newblock {\em Zeitschrift f{\"u}r Physikalische Chemie} 216(2):559.

\bibitem{Moler_2003}
Moler C, Van~Loan C (2003) {Nineteen Dubious Ways to Compute the Exponential of
  a Matrix, Twenty-Five Years Later}.
\newblock {\em SIAM Rev.} 45(1):3--49.

\bibitem{Smerlak_2019}
Smerlak M (2019) {Effective potential reveals evolutionary trajectories in
  complex fitness landscapes}.
\newblock arXiv:1912.05890.

\bibitem{Szendro_2013}
Szendro IG, Schenk MF, Franke J, Krug J, de~Visser JAGM (2013) {Quantitative
  analyses of empirical fitness landscapes}.
\newblock {\em J. Stat. Mech.} 2013(01):P01005.

\bibitem{L_ssig_2017}
L{\"a}ssig M, Mustonen V, Walczak AM (2017) {Predicting evolution}.
\newblock {\em Nature Ecology {\&} Evolution 2017 1:3} 1(3):20150057.

\end{thebibliography}

\end{document}